\documentclass [11pt,letterpaper]{JHEP3}
\usepackage{cite}	
\usepackage{bm}		
\usepackage{verbatim}   
\advance\parskip 1.0pt plus 1.0pt minus 2.0pt	
\def\coeff#1#2{{\textstyle {\frac {#1}{#2}}}}
\def\half{\coeff 12}
\def\l{\ell}
\def\j{{\bf j}}
\def\g{{\bf g}}
\def\h{{\bf h}}
\def\r{{\bf r}}
\def\P{{\bf P}}
\def\G{{\bf G}}
\def\d{{\rm d}}
\def\tr{{\rm tr}}

\def\Nc{N_{\rm c}}
\def\Ns{N_{\rm s}}
\def\Nf{N_{\rm f}}
\def\Z{{\mathbb Z}}

\def\parent{{\rm (p)}}
\def\daughter{{\rm (d)}}

\def\U{{\hat U}}
\def\E{{\hat E}}
\def\phih{{\hat \phi}}
\def\pih{{\hat \pi}}
\def\psih{{\hat \psi}}
\def\H{{\hat H}}

\def\T{{\bf T}}
\def\betatilde{\widetilde\beta}

\preprint {UW/PT 04--23\\INT-PUB 04--31}
\title
    {%
    Necessary and sufficient conditions for non-perturbative equivalences
    of large $\bm \Nc$ orbifold gauge theories
    }%
\author
    {%
    Pavel Kovtun,\footnote
	{%
	Current address: KITP, University of California,
	Santa Barbara, CA 93106
	}~
    Mithat \"Unsal,\footnote
	{%
	Current address: Department of Physics,
	Boston University,
	Boston, MA 02215
	}~
    and Laurence G.~Yaffe
    \\Department of Physics
    \\University of Washington
    \\Seattle, Washington 98195--1560
    \\Email:
    \parbox[t]{2in}{\email {pkovtun@phys.washington.edu},\\
		    \email {mithat@phys.washington.edu},\\
		    \email {yaffe@phys.washington.edu}}
    }%

\abstract
    {%
    Large $N$ coherent state methods are used to study the relation
    between $U(\Nc)$ gauge theories containing adjoint representation
    matter fields and their orbifold projections.
    The classical dynamical systems
    which reproduce the large $\Nc$ limits of the quantum dynamics
    in parent and daughter orbifold theories are compared.
    We demonstrate that the large $\Nc$ dynamics of the parent theory,
    restricted to the subspace invariant under the orbifold projection
    symmetry,
    and the large $\Nc$ dynamics of the daughter theory,
    restricted to the untwisted sector invariant under ``theory space''
    permutations, coincide.
    This implies equality, in the large $\Nc$ limit, between appropriately
    identified connected correlation functions in parent and daughter theories,
    provided the orbifold projection symmetry is not spontaneously broken
    in the parent theory and the theory space permutation symmetry is not
    spontaneously broken in the daughter.
    The necessity of these symmetry realization conditions
    for the validity of the large $\Nc$ equivalence is unsurprising,
    but demonstrating the sufficiency of these conditions is new.
    This work extends an earlier proof of non-perturbative large $\Nc$
    equivalence which was only valid in the phase of the (lattice regularized)
    theories continuously connected to large mass and
    strong coupling \cite{KUY}.
    }%

\keywords{1/N Expansion, Lattice Gauge Field Theories}

\begin {document}

\setlength{\baselineskip}{1.10\baselineskip}

\section {Introduction}

This paper is devoted to a comparison of the dynamics of
large $\Nc$ gauge theories related by orbifold projections.
In this context, orbifold projection is a technique
for constructing ``daughter'' theories starting from some
``parent'' theory, by retaining only those fields which
are invariant under a chosen discrete symmetry group
of the parent theory.
In suitable cases,
planar graphs of the daughter theory coincide with
the planar graphs of the original theory, up to
a simple rescaling of the gauge
coupling constant~\cite{Bershadsky-Johansen}.
This implies that the large $\Nc$ limits of the parent
and daughter theories have coinciding perturbative expansions.
Previous work
\cite{Schmaltz,Erlich-Naqvi,Strassler,
Gorsky-Shifman,Dijkgraaf-Neitzke-Vafa,Tong}
has examined various tests in an effort to determine
whether these large $\Nc$ equivalences hold non-perturbatively,
and has explored interesting consequences which would follow
from such non-perturbative equivalences.
Examples were found where a non-perturbative
equivalence appears to hold, and also where it fails, but a clear
delineation of the domain of validity of large $\Nc$ equivalences
between parent and daughter orbifold gauge theories has not yet been given.

Recently, a rigorous proof of large $\Nc$ parent/daughter equivalence has been
constructed for a large class of Euclidean lattice gauge theories,
in the phase of both theories which is continuously connected
to strong gauge coupling and large mass~\cite{KUY}.
This proof relied on the observations that
(\emph{i}) loop equations for relevant gauge-invariant observables
coincide in both theories,%
\footnote{
	The loop equations coincide provided
	the 't Hooft couplings of the two theories are equal
	and provided certain symmetries are not
	spontaneously broken.
	These symmetry realization conditions are discussed below.
	}
and
(\emph{ii})
the set of loop equations uniquely determine the corresponding
correlation functions in the strong coupling, large mass phase of the theory.
Coinciding loop equations are not, by themselves, sufficient
to prove equivalence because there may be (and typically are) multiple
solutions to the infinite set of loop equations.
Hence, parent and daughter theories might correspond to {\em different}
solutions of the same set of loop equations.
The convergence of the strong coupling and small hopping parameter
expansion was used to rule out this possibility in Ref.~\cite {KUY},
but this restricted the resulting proof of equivalence to the phase
of the lattice theories smoothly connected to strong coupling and large mass.

The goal of this paper is to identify necessary and sufficient
conditions for the validity of large $\Nc$ equivalence
in a wide class of orbifold projections,
in a form which is
applicable to any phase of the lattice-regulated theories.
Our basic strategy will involve constructing, and comparing,
the classical dynamical systems which reproduce the large $\Nc$
quantum dynamics of the parent and daughter theories \cite{LGY-largeN}.%
\footnote{
	For ease of presentation, we will choose to work
	with Hamiltonian lattice gauge theories.
	One could instead use Euclidean lattice formulations
	and compare the corresponding large $\Nc$ coherent state
	free energies \cite{LGY-largeN2,LGY-largeN3}.
	}
This requires identifying the appropriate
infinite dimensional group which generates
gauge invariant large $\Nc$ coherent states
in gauge theories with matter fields.
The resulting large $\Nc$ classical dynamics contains all the information
needed to reconstruct connected correlation functions in the
corresponding quantum theory in the $\Nc\to\infty$ limit.
More physically, this means the leading large $\Nc$ behavior
of meson or glueball masses, decay widths, or scattering amplitudes
may be extracted from the large $\Nc$ classical dynamics.

We will argue that comparison of
large $\Nc$ classical Hamiltonians (and corresponding phase spaces)
provides a sufficient means for determining when two theories have
coinciding large $\Nc$ limits.
This description of the large $\Nc$ dynamics is valid in any phase
of the theory.
For any given choice of coupling constants, the minimum of the classical
Hamiltonian determines the correct ground state.
The large $\Nc$ loop equations are equivalent to stationarity conditions
for the large $\Nc$ classical Hamiltonian.
The difficulty with multiple solutions of the loop equations is avoided
by working with the Hamiltonian directly, since its global minimum
identifies the correct solution to the loop equations.

The paper is organized as follows.
Section 2 establishes our notation and
briefly reviews the construction of large $\Nc$
classical dynamics.
A key ingredient will be the construction of an infinite dimensional Lie group,
termed the {\em coherence group}, which generates suitable coherent states.
The classical phase space is a coadjoint orbit
(or a particular coset space)
of the coherence group.
In Section~3
we apply the formalism to $U(\Nc)$ gauge theories
with adjoint matter fields,
while Section 4 applies the coherent state formalism
to orbifold projections of $U(\Nc)$ gauge theories
with adjoint matter fields.
We examine the large $\Nc$ classical dynamics in the sector of the
parent theory which is invariant under the symmetry used to impose
the orbifold projection, and in the sector of the daughter theory
which is invariant under the global symmetry which
permutes equivalent gauge group factors.
Section 5 explains how the equivalent structure of the coherence algebras,
when restricted to these symmetry sectors, implies equivalent
classical Hamiltonian descriptions.

\section{Large $\bm N$ limits as classical mechanics}
\label{sec:method}

The key observation for the classical description of
large-$N$ theories is the factorization of expectation values
of products of operators in the $N\to\infty$ limit.
This is analogous to the suppression of quantum fluctuations,
and consequent factorization of expectation values,
in the small-$\hbar$ limit of point particle quantum mechanics.

One way to understand the emergence of classical mechanics
in the $\hbar\to 0$ limit of quantum mechanics
is via the properties of coherent states.
In point particle quantum mechanics, these are states whose
uncertainties in both position and momentum vanish as $\hbar\to 0$.
The set of coherent states provides an overcomplete basis,
and the overlap between two different coherent states vanishes
exponentially (in $1/\hbar$) as $\hbar \to 0$.
Classical observables are the small $\hbar$ limits of coherent state
expectation values of the corresponding quantum operators,
and Poisson brackets are coherent state expectation
values of the corresponding commutators
(properly scaled to have a finite limit as $\hbar\to0$).
The set of coherent states can be constructed by the
action of the Heisenberg group
(the group of translations in position or momentum) applied to
some initial base state, for example a state with a Gaussian wave function.

Exactly the same strategy may be employed to
construct the classical dynamics which reproduces the large $N$ limit
of quantum dynamics of theories
such as $O(N)$ invariant spin models, or $U(N)$ gauge theories
\cite{LGY-largeN}.
The analog of the Heisenberg group is
a \emph{coherence group} $\mathbf{G}$ which is represented
by a set of unitary operators
$\{\hat\mathbf{G}(u)\}$,
$u\in\mathbf{G}$.
The Lie algebra $\mathbf{g}$ for the coherence group
(or the \emph{coherence algebra})
consists of generators which are represented
by anti-Hermitian operators
$\{\hat\Lambda(\alpha)\}$, $\alpha\in\mathbf{g}$.
Generalized coherent states are generated by the
action of the coherence group on some
base state $|0\rangle$,
\begin{equation}
     |u\rangle = \hat\mathbf{G}(u) |0\rangle\,.
\end{equation}
The phase space and Poisson brackets of the resulting classical
dynamical system are completely determined by the structure
of the coherence group.%
\footnote{
	Phase space may be identified with a coadjoint orbit
	of the coherence group, and the Poisson bracket is given by the
	Kirillov form \cite{Arnold,LGY-largeN}.
	The details of this construction will not be essential
	for our purpose.
	The conditions which a valid coherence group must satisfy
	are discussed in the next section and more thoroughly
	in Ref.~\cite{LGY-largeN};
	these are crafted so as to ensure the validity of the
	basic results (\ref {eq:A})--(\ref {eq:[A,B]}) below.
	}
For $U(\Nc)$ gauge theories, the coherence algebra $\mathbf{g}$ will be
the set of all anti-Hermitian linear combinations of spatial Wilson loops,
plus loops decorated with electric field or matter field insertions.
Correspondingly, expectation values of decorated Wilson loops
may be thought of as providing coordinates on phase space.

Operators whose coherent state matrix elements
$\langle u|\hat A|u'\rangle/\langle u|u'\rangle$
are finite 
in the limit $\Nc\to\infty$
will be called \emph{classical}.
The overlap of two different coherent states, as well as
off-diagonal coherent state matrix elements of classical
operators, decay exponentially as $\Nc\to\infty$.
The key relations between functions on the phase space
(classical observables) and matrix elements of classical operators are:
\begin{equation}
   \lim_{\Nc\to\infty} \langle u|\hat A|u\rangle=a(\zeta) \,,
\label{eq:A}
\end{equation}
\begin{equation}
   \lim_{\Nc\to\infty} \langle u|\hat A \hat B|u\rangle=a(\zeta) \, b(\zeta) \,,
\label{eq:AB}
\end{equation}
\begin{equation}
   \lim_{\Nc\to\infty} i\Nc^2 \, \langle u|[\hat A,\hat B]|u\rangle=
        \{a(\zeta),b(\zeta)\}_{PB} \,.
\label{eq:[A,B]}
\end{equation}
Here $\zeta$ denotes a point in phase space, and the first
relation is the definition of the classical observable $a(\zeta)$
corresponding to an operator $\hat A$.
The second relation is a statement of
factorization in the large-$\Nc$ limit.
The third relation ensures that quantum dynamics
turns into classical dynamics,
${d a(\zeta)}/{dt} = \{h_{\rm cl}(\zeta), a(\zeta) \}_{PB}$,
where the large-$\Nc$ classical Hamiltonian is
\begin{equation}
   h_{\rm cl}(\zeta) \equiv \lim_{\Nc\to\infty} \frac{1}{\Nc^2} \>
                  \langle u|\H|u\rangle\,.
\label{eq:hcl}
\end{equation}

Finding the ground state of the original quantum theory reduces,
in the $\Nc\to\infty$ limit,
to locating the point $\zeta_{\rm min}$ in the classical
phase space which minimizes the large $\Nc$ classical Hamiltonian $h_{\rm cl}$.
Coherent state expectation values evaluated at $\zeta_{\rm min}$
give the large $\Nc$ limits of ground state expectation values in the
original quantum theory.
Linearizing the classical equations of motion about $\zeta_{\rm min}$
and solving for the resulting small oscillation frequencies directly
yields the large $\Nc$ limit of excitation energies to low-lying
excited states.
The leading large $\Nc$ behavior of decay widths and scattering amplitudes
are determined by higher derivatives of the classical action evaluated
at $\zeta_{\rm min}$.
Similarly, 
connected correlation functions of products of $K$ classical operators,
when multiplied by $(\Nc^2)^{K-1}$,
have non-trivial large $\Nc$ limits which are determined by
derivatives up to order $K$ of the classical action
\cite{LGY-largeN}.

The large-$N$ limits of two different quantum theories will be identical
if the classical dynamical systems generated by these
theories are identical; in other words, if both phase
spaces and classical Hamiltonians can be appropriately identified.
In what follows we will not actually compare the full large-$N$ classical
dynamical systems of two theories,
but rather the large-$N$ dynamics in specific sectors of
each theory which are invariant under certain symmetries.
The coherent state construction implies that the
large-$N$ dynamics of corresponding sectors of two theories are
identical if
(\emph{i}) the coherence subgroups
(which leave invariant the chosen sectors) are isomorphic,
(\emph{ii}) the action of the coherence subgroups on corresponding
observables is isomorphic,
and (\emph{iii}) the base state expectation values of corresponding
observables coincide.
Together, these conditions ensure
isomorphism between the chosen sectors of phase space in each theory,
as well as the proper identification of corresponding
observables in the two theories.

\section {Parent theory}

\subsection{Gauge theories with adjoint matter fields}

To establish notation,
we first briefly review the Hamiltonian
formulation of lattice gauge theories
containing matter fields in the adjoint representation
of the gauge group.
(For a lengthier introduction see, for example, Ref.~\cite{Creutz}.)
In a $U(\Nc)$ gauge theory,
the gauge field degrees of freedom
are link variables (coordinates) $\U_{ij}[\l]$, and
electric field operators (conjugate momenta) $\E_{ij}[\l]$,
associated with oriented links $\l$ of
a spatial lattice $\Lambda$
(of any dimensionality).
For simplicity, $\Lambda$ will
be assumed to be a simple cubic lattice.
The $\Nc\times\Nc$ matrix $\U[\l] \equiv \| \U_{ij}[\l] \|$ is unitary,
while $\E[\l] \equiv \| \E_{ij}[\l] \|$ is Hermitian.
We will use
$\bar\l$ to denote the oppositely directed link to $\l$,
with $\U[\bar\l] \equiv \U[\l]^\dagger = (\U[\l])^{-1}$,
and $\E[\bar\l] \equiv -\U[\bar\l] \E[\l] \U[\l]$.%
\footnote
    {
    The dagger acts both on quantum operators
    and on the matrix indices, so
    $(\hat U[\l]^\dagger)_{ij} \equiv (\hat U[\l]_{ji})^\dagger$, {\em etc}.
    }
These operators obey the
canonical commutation relations:
\begin{eqnarray}
   \label{eq:commPU}
   \left[ \U_{ij}[\l], \U_{kl}[\l'] \right] &=& 0, \\
   \left[ \E_{ij}[\l], \U_{kl}[\l'] \right] &=&
         \frac{1}{\Nc} \, \delta_{\l \l'} \, \delta_{kj} \, \U_{il}[\l], \\
   \left[ \E_{ij}[\l], \U_{kl}[\bar{\l'}] \right] &=&
         -\frac{1}{\Nc} \, \delta_{\l \l'} \, \delta_{il} \, \U_{kj}[\bar\l], \\
   \left[ \E_{ij}[\l], \E_{kl}[\l'] \right] &=&
         \frac{1}{\Nc} \, \delta_{\l \l'} \left(\delta_{kj} \E_{il}[\l]
         - \delta_{il} \E_{kj}[\l]\right) \,,
\end{eqnarray}
with the gauge indices $i,j$, {\em etc.} running from $1$ to $\Nc$.

Matter fields in the adjoint representation
of the gauge group may be added
by placing canonically conjugate pairs
of operators at the sites of the lattice.
Specifically, at each site $s\!\in\!\Lambda$
we add $\Ns$ complex scalars
$\phih^{a}_{ij}[s]$
and their conjugate momenta
$\pih^{a}_{ij}[s]$
(with $a=1,\dots,\Ns$),
as well as $\Nf$ fermion fields
$\psih^{b}_{ij}[s]$
($b=1,\dots,\Nf$).
These matter field satisfy the following
(anti-)commutation relations:
\begin{eqnarray}
    \left[ \phih^{a}_{ij}[s], \pih^{a'}_{lk}[s']^\dagger \right]
    &=&
    \left[ \phih^{a}_{ij}[s]^\dagger, \pih^{ a'}_{lk}[s'] \right]
    =
    \frac{i}{\Nc}\, \delta_{ss'}\, \delta^{a a'}\, \delta_{il}\, \delta_{kj} \,,
\label{eq:comm-rel-scalar}
\\
    \left\{ \psih^{b}_{ij}[s], \psih^{b'}_{lk}[s']^\dagger \right\}
    &=&
    \frac{1}{\Nc}\, \delta_{ss'}\, \delta^{b b'}\, \delta_{il}\, \delta_{kj} \,.
\label{eq:comm-rel-fermion}
\label{eq:commPpsi}
\end{eqnarray}
The Hamiltonian of the ``parent'' theory will be taken to be
\begin{equation}
  \H^\parent \equiv
  \H^\parent_{\rm gauge} +
  \H^\parent_{\rm scalar} +
  \H^\parent_{\rm fermion}\,,
\label {eq:parentHamiltonian}
\end{equation}
with $\H^\parent_{\rm gauge}$ the standard Kogut-Susskind
Hamiltonian~\cite{Kogut-Susskind},
\begin{equation}
   \H^\parent_{\rm gauge} =
	\Nc \, \biggl\{\,
	\frac{1}{4\betatilde} \sum_{\l\in\Lambda} \> \tr\, \E[\l]^2 -
	\betatilde \, \sum_{p\in\Lambda} \>
	\tr\Bigl( \U[\partial p] + \U[\overline{\partial p}] \Bigr)
	\biggr\}\,,
\label {eq:pHgauge}
\end{equation}
with the inverse 't Hooft coupling
$\betatilde\equiv\beta/\Nc = 1/(g^2 \Nc)$
held fixed as $\Nc{\to}\infty$.
(Throughout this paper, we set the lattice spacing to one.)
The plaquette variable $\hat U[\partial p]$ denotes the ordered
product of link variables around the boundary of the plaquette $p$,
and $\overline{\partial p}$ is the oppositely oriented plaquette boundary.
More generally, $\hat U[C]$ will denote the product of link variables
around an arbitrary closed loop $C$.
The scalar field Hamiltonian will be defined as
\begin{eqnarray}
   \H^\parent_{\rm scalar} &=&
    \Nc \,
    \sum_{s\in \Lambda} \;
    \Bigl\{
	\tr \, \Bigl(\pih^a[s]^\dagger \, \pih^a[s]\Bigr)
    +
    \Nc\, V\!\Bigl[
	\tr \, \Bigl(\phih^a[s]^\dagger\,\phih^a[s] \Bigr)/\Nc
    \Bigr]
    \Bigr\}
\nonumber\\
    &-&
    \Nc
    \sum_{\l=\langle ss' \rangle \in\Lambda}
	\frac \kappa 2 \>
	\tr \, \Bigl(
	\phih^a[s]^\dagger \, \U[\l] \, \phih^a[s'] \, \U[\bar \l]
	\Bigr) \,,
\label {eq:pHscalar}
\end{eqnarray}
while the fermion Hamiltonian is
\begin{equation}
   \H^\parent_{\rm fermion} =
     \Nc \, \biggl\{
     \frac \kappa {2i}
     \sum_{\l=\langle ss' \rangle \in\Lambda}\kern-5pt
	\tr \,
	\Bigl(
	    \psih^b[s]^\dagger \, \eta[\l] \, \U[\l] \, \psih^b[s'] \,
	    \U[\bar \l]
	\Bigr)
    +
    m_{\rm f} \sum_{s\in\Lambda} \> \xi_s\,
    \tr \, \Bigl(\psih^b[s]^\dagger \, \psih^b[s]\Bigr)
    \biggr\}\, .
\label {eq:pHfermion}
\end{equation}
Sums over repeated flavor indices are implied.
The total Hamiltonian has been constructed to be invariant under
a $U(\Ns) \times U(\Nf)$ flavor symmetry,
in addition to the local $U(\Nc)$ gauge symmetry.
The ``hopping parameter'' $\kappa$ has been chosen,
without loss of generality, to be the same for scalars and fermions.
The exact shape of the scalar potential
$V[\chi]$ will not be important.
We have chosen to use a ``staggered'' discretization of
fermion fields \cite {staggered}.
In the fermion Hamiltonian (\ref{eq:pHfermion}),
$\eta[\l]$ is an
imaginary phase factor assigned to each link
in such a way that the product of these phases
around every plaquette is minus one,
$\eta[\partial p]=-1$.
(For links with reversed orientation, $\eta[\bar\l]\equiv\eta[\l]^*$.)
The factor $\xi_s$ in the mass term is a sign factor
which equals $+1$ for sites on the even sublattice
of $\Lambda$ and $-1$ for sites on the odd sublattice.%
\footnote
    {
    The factor $\xi_s$ plays the role, for staggered fermions,
    that $\gamma_0$ normally does.
    With this factor, the mass term is minimized for half filling;
    without this factor, $\sum_s \tr ( \psih^b[s]^\dagger \psih^b[s])$
    would be the conserved fermion number, not a mass term.
    }

\subsection{Coherence algebra and coherent states}

The coherence algebra in a $U(\Nc)$ pure gauge theory
may be taken to be the set of all anti-Hermitian
linear combinations of Wilson loops and
loops with one electric field insertion,
\begin{equation}
     \hat\Lambda_{\rm gauge}(a,b) \equiv
		 \Nc \> \biggl\{ \>
                 \sum_{C\subset\Lambda} \> a^C \,\tr (\U[C])
               + \sum_{C\subset\Lambda} \sum_{\l\subset C} \>
                 b^{\l C} \, \tr({:}\E[\l]\,\U[C]{:})
		 \biggr\} \,,
\label {eq: gauge-algebra}
\end{equation}
where normal ordering is defined as
\begin{equation}
     {:}\E[\l]\,\U[C]{:} \equiv
     \half \left(\E[\l]\,\U[C]+\U[C]\E[\l] \right) ,
\end{equation}
and $C$ denotes a closed loop in $\Lambda$
(beginning with link $\l$ or ending with $\bar\l$).
The overall factor of $\Nc$ is included so that the
structure constants of the coherence algebra are $\Nc$-independent,
given our chosen normalization for the canonical commutation relations.

For gauge theories containing adjoint matter fields,
one must enlarge the coherence algebra to include Wilson loops
with matter field insertions.
Writing explicit expressions for such decorated loops can be tedious,
but may be avoided if one introduces a higher-dimensional
``extended'' lattice in which links pointing in new directions
represent insertions of adjoint representation matter fields.
This is discussed in detail in Ref.~\cite{KUY}.
For theories containing fermions, the result is an extended lattice
$\bar\Lambda$ whose sites are two copies of the sites of $\Lambda$.
That is, for every site $s \in \Lambda$, one introduces a $\Z_2$ partner
$s'$.
Emanating from every site are $\Ns$ oriented ``scalar links'' which
return to the same site, plus $\Nf$ oriented ``fermion links'' which
connect the site with its $\Z_2$ partner site,
in addition to the gauge links of the original lattice.%
\footnote
    {
    This is the appropriate extended lattice for bosonic operators
    (containing an even number of fermions).
    For fermionic operators, there is no need for this doubling of sites,
    and fermionic links may be regarded as returning to the same site
    from which they originate.
    This distinction reflects the fact that moving a fermion
    from one end of a trace to the other involves an odd number
    of interchanges with other fermion operators if the overall
    trace is bosonic, but an even number if the trace is fermionic.
    For simplicity, we will not bother to distinguish explicitly
    the bosonic and fermionic extended lattices in the following discussion.
    }
Every gauge-invariant decorated Wilson loop on the original lattice
may be represented by a closed loop on the extended lattice ---
and vice versa.
It will also be convenient to define $\bar\Lambda_{\rm s}$ as the subset
of the extended lattice $\bar\Lambda$ which omits the $\Z_2$ partner sites
and all the fermion links,
so that loops in $\bar\Lambda_{\rm s}$ are Wilson loops decorated with
arbitrary scalar insertions.

Using this representation,
an appropriate coherence algebra
for $U(\Nc)$ gauge theories containing
adjoint matter,
which we will denote as $\mathbf{g}_{\rm parent}$,
consists of all anti-Hermitian operators of the form%
\footnote
    {%
    For loops containing scalar conjugate momenta,
    $
	{:}\pih^a[s] \hat U[C]{:}
	\equiv
	\half ( \pih^a[s] \hat U[C] + \hat U[C] \pih^a[s] )
    $.
    For loops with two fermion insertions, one may similarly define
    $
	{:}\psih^b[s] \hat U[\Gamma_1] \psih^{b'}[s'] \hat U[\Gamma_2] {:}
	\equiv
	\half (
	\psih^b[s] \hat U[\Gamma_1] \psih^{b'}[s'] \hat U[\Gamma_2]
	-
	\psih^{b'}[s'] \hat U[\Gamma_2] \psih^b[s] \hat U[\Gamma_1]
	)
    $
    and have a closed sub-algebra.
    For loops with more than two fermion insertions, one must allow
    multiple operator orderings in order to have a closed algebra.
    Hence, for such loops
    the coefficients $a^C$ in the algebra (\ref {eq:gsf-algebra})
    really depend on a specification of operator ordering in addition
    to the choice of a geometric loop $C$ on the extended lattice.
    This is not a significant issue for our purposes,
    and will not be indicated explicitly.
    For fermionic loops
    ({\em i.e.}, loops containing an odd number of fermion insertions),
    the corresponding coefficient must be understood as a generator of a
    Grassmann algebra, not a $c$-number.
    [Alternatively, one may omit all fermionic loops
    and restrict attention to the bosonic sector of the theory.]
    }
\begin{eqnarray}
    \hat\Lambda^\parent(a,b,c)
    &\equiv &
    \Nc \, \biggl\{
      \sum_{C\subset\bar\Lambda} a^C \, \tr (\U[C])
    + \sum_{C\subset\Lambda} \sum_{\l\subset C} \>
	b^{\l C} \, \tr({:}\E[\l]\,\U[C]{:})
\nonumber\\
    &+&
    \sum_{C\subset\bar\Lambda_{\rm s}} \, \sum_{s\subset C} \>
    \left[ \,
	c_a^{s C} \, \tr({:}\pih^a[s]\,\U[C]{:})
	- (c_a^{s C})^* \, \tr ({:}\pih^a[s]^\dagger\,\U[\bar C]{:})
    \right]
    \biggr\} \,.
    \label {eq:gsf-algebra}
\end{eqnarray}
More descriptively, the algebra consists of
arbitrary anti-Hermitian linear combinations of Wilson
loops containing arbitrary scalar and fermion field insertions,
loops containing arbitrary scalar field insertions plus
one scalar conjugate momentum insertion,
and ordinary Wilson loops with one electric field insertion.
One can easily see that the above coherence algebra is closed,
and that the coherence algebra for pure gauge theory
forms a normal subalgebra.
The coherence group $\G_{\rm parent}$
generated by the algebra (\ref{eq:gsf-algebra})
acts irreducibly on
the Hilbert space of the theory.%
\footnote
    {%
    This means that the only operator which commutes
    with all elements of the group is proportional to the identity.
    This is most easily verified using the coordinate space representation
    where link variables and scalar fields are diagonal,
    and conjugate momenta act as derivatives.
    }
This implies that the coherent states
$ |u \rangle \equiv \hat \G(u) |0\rangle$
generated by elements of this group acting on any initial state
$|0\rangle$ are (over)-complete.

We will choose the base state
to be a simple product state,
\begin {equation}
    |0\rangle =
    |0\rangle_{\rm gauge} \otimes
    |0\rangle_{\rm scalar} \otimes
    |0\rangle_{\rm fermion} \,,
\end {equation}
where $|0\rangle_{\rm gauge}$ is
the infinite-coupling pure gauge ground state,
and $|0\rangle_{\rm scalar}$ is a decoupled product
of Gaussian states
annihilated by $\phih^a[s] + i \pih^a[s]$ and
$\phih^a[s]^\dagger +i \pih^a[s]^\dagger$.
Coordinate representation wavefunctions for these states are
\begin{eqnarray}
    &&
    \langle U|0\rangle_{\rm gauge} = 1 \,,
\label{eq:gauge-wf}
\qquad
    \langle \phi|0\rangle_{\rm scalar} =
      \prod_{s\in\Lambda}
      \prod_a \>
      e^{-\Nc \, \tr (\phi^a[s] \phi^a[s]^*)}\,.
\label{eq:scalar-wf}
\end{eqnarray}
The fermion state $|0\rangle_{\rm fermion}$ will be taken to be the
ground state of the fermionic mass term in (\ref {eq:pHfermion}),
or the state which is annihilated by $\psih^b[s]$ on all even sites and by
$\psih^b[s]^\dagger$ on all odd sites.
Hence, it has $\Nc^2 \, \Nf$ fermions at every site of the odd sublattice,
and no fermions at even sublattice sites.
The resulting product state is fully gauge-invariant.
The fact that the wave function (\ref{eq:gauge-wf}) has no dependence
on link variables
implies that the base state expectation value of
any decorated Wilson loop which is not local to a single site
vanishes in the $\Nc\to\infty$ limit.
The only observables which have non-zero
base state expectation values at $\Nc=\infty$
are traces of products of matter fields at a single site.

Using arguments analogous to those presented in Ref.~\cite{LGY-largeN},
one can also show that no operator (except zero) has
identically vanishing expectation values in the coherent states generated by
$\mathbf{G}_{\rm parent}$ acting on this base state.
This implies that any operator is uniquely defined by its
diagonal coherent state expectation values.
This result, plus the irreducible action of $\G_{\rm parent}$
on the Hilbert space,
are the key conditions needed to show that quantum dynamics
reduces, in the $\Nc\to\infty$ limit, to classical dynamics on a phase space
which is a coadjoint orbit of the coherence group $\G_{\rm parent}$,
with a classical
Hamiltonian given by the large $\Nc$ limit of the coherent state
expectation value of the quantum Hamiltonian (suitably scaled)
\cite {LGY-largeN},
\begin{equation}
   h^\parent_{\rm cl}(\zeta) \equiv \lim_{\Nc\to\infty} \frac{1}{\Nc^2} \>
                  \langle u|\H^\parent|u\rangle\,.
\label{eq:phcl}
\end{equation}

\subsection {Symmetry invariant dynamics}

The lattice theory defined by the parent Hamiltonian
(\ref{eq:parentHamiltonian}) has a
$U(\Nc)\times G_{\rm flavor}$ global symmetry group,
where the $U(\Nc)$ factor represents space
independent ({\em i.e.}, site independent) gauge transformations,
and $G_{\rm flavor} \equiv U(\Ns){\times}U(\Nf)$ represents flavor rotations
of the scalars and fermions.
To define a daughter theory via an orbifold projection,
one chooses a discrete ``projection group'' \P\
which is a subgroup of this global symmetry group
and eliminates from the theory all degrees of freedom except those
invariant under the chosen group \P.
Specifics of this choice, and the resulting form of daughter theories,
will be discussed in the next section.

Gauge invariant operators which commute with the chosen projection
group \P\ will be termed {\em neutral}.
Neutral elements of the coherence group $\G_{\rm parent}$
necessarily form a subgroup,
that we will call ${\bf H}_{\rm parent}$, which is generated by
the neutral subalgebra $\h_{\rm parent}$ of the coherence algebra
$\g_{\rm parent}$.
Our chosen base state $|0\rangle$ is invariant under both gauge
transformations and flavor rotations,
and thus will be invariant under the projection group \P\
(whatever its choice).
Therefore, elements of the neutral coherence subgroup ${\bf H}_{\rm parent}$
acting on the base state will generate \P-invariant coherent states,
which provide an over-complete basis for the \P-invariant
sector of the Hilbert space.
Everything said above regarding the construction of large $\Nc$ classical
dynamics may be specialized to the \P-invariant sector of the theory.
Large $\Nc$ quantum dynamics in this sector is equivalent to classical
dynamics on the \P-invariant subspace of the full large $\Nc$ phase space,
which is a coadjoint orbit of the neutral coherence subgroup
${\bf H}_{\rm parent}$.

There is one important caveat associated with restricting attention
to the \P-invariant sector of the large $N$ phase space:
whether dynamics in this sector is particularly {\em interesting}
depends on the symmetry realization of $\P$.
If this symmetry is not spontaneously broken,
then the minimum of the classical Hamiltonian will lie in the
\P-invariant subspace of the phase space.
Hence, minimizing the Hamiltonian or studying small oscillation
frequencies within this subspace will yield information about
ground state properties or low energy excitations in the underlying
quantum theory.
On the other hand, spontaneous breaking of the symmetry \P\ means
that the minimum of the classical Hamiltonian does not lie
on the $\P$-invariant subspace of the phase space, but rather
that there are multiple degenerate minima related by the symmetry.
In this case, understanding classical dynamics within the \P-invariant
subspace will not teach one anything about ground state properties
of the underlying quantum theory.%
\footnote
    {%
    Some readers may object that even if there is
    spontaneous symmetry breaking,
    and hence degenerate ground states,
    one can always construct a \P-invariant ground state
    as a superposition of the non-invariant ground states.
    However, the quantum dynamics in this \P-invariant state does not reduce,
    in the large $\Nc$ limit, to classical dynamics of a pure
    state in the \P-invariant sector of phase space.
    Rather, the large $\Nc$ dynamics of this state is indistinguishable
    from the classical dynamics of a mixed state which is a statistical
    average of the non-invariant minima of the classical Hamiltonian.
    This reflects the fact that states which violate cluster decomposition
    also necessarily violate large $N$ factorization;
    such states are indistinguishable from mixed states in the large $N$
    limit.
    }

\section{Daughter theories}

\subsection {Projection group}

We will limit our consideration to Abelian projection groups.
The maximal Abelian subgroup of the flavor symmetry group is $U(1)^p$,
where $p\equiv \Ns+\Nf$.
We will pick the projection group to be a product of $p$ cyclic groups
with orders $k_1,\ldots,k_p$, so
\begin{equation}
     \P
     =
     \mathbb{Z}_{k_1}\times\mathbb{Z}_{k_2}\times\cdots\times\mathbb{Z}_{k_p}
     \,.
\end{equation}
Let $m \equiv k_1 \, k_2 \cdots k_p$ denote the dimension of this group.
We require $\Nc$ to be divisible by $m$,
so that $\Nc = m \, N$ for some integer $N$.

Interesting daughter theories,
with gauge groups differing from their parent,
result from choosing a projection group which involves
global gauge transformations in addition to possible flavor rotations.
The embedding of $\P$ within the global symmetry group will be chosen
so that only a $[U(N)]^{m}$ subgroup of the original
$U(\Nc)$ gauge group commutes with $\P$.
This means that the resulting daughter theory will have a
product gauge group consisting of $m$ separate
$U(N)$ factors.
The generator $\eta_j$ of each cyclic group $\mathbb{Z}_{k_j}$
will be chosen to be the product of a gauge transformation
times a flavor transformation
lying in the maximal Abelian subgroup,
$\eta_j = \gamma_j \times \zeta_j$ with
$\gamma_j \in U(\Nc)$ and $\zeta_j \in G_{\rm flavor}$.
The gauge transformations $\{ \gamma_j \}$ may be chosen as
\begin {equation}
    \gamma_j
    =
    \underbrace{1_{k_1} \times\cdots}_{j-1}{}\times \Omega_j \times
    \underbrace{1_{k_{j+1}}\times\cdots}_{p-j}{} \times 1_N \,,\ \ \ \
    j=1,\ldots,p
\end {equation}
where
$1_{k_j}$ denotes a $k_j\times k_j$ identity matrix, and
$\Omega_j$ is the diagonal matrix of size $k_j$
whose diagonal elements are all $k_j$'th roots of unity,
$\Omega_j \equiv {\rm diag} (1,\omega_j,\cdots,\omega_j^{k_j-1})$
with
$\omega_j=e^{2\pi i/k_j}$.
The associated flavor rotations $\{ \zeta_j \}$ may be written
\begin {equation}
    \zeta_j = e^{2\pi i \hat r_j/k_j} \,,
\end {equation}
where each $\hat r_j$ is a charge operator which assigns integer values,
defined modulo $k_j$,
to matter fields in the theory
(and zero to all links variables and electric field operators).
The entire set of charge operators may be regarded as
a $p$-component vector, $\hat \r \equiv \{ \hat r_j \}$.
The $\r$-charge assignments of each scalar and fermion field,
{\em i.e.}\ $\{ \r[\phih^a] \}$ and $\{ \r[\psih^b] \}$,
may be chosen arbitrarily.
Different choices will lead to different daughter theories.
The canonical commutation relations (and Hermiticity) imply that
$
    \r[\pih^a]
    = \r[\phih^a]
    = -\r[\phih^a{}^{\dagger}]
    = -\r[{\pih^a}{}^\dagger]
$,
and
$\r[\psih^b] = -\r[\psih^b{}^\dagger]$.
The $\r$-charge of an operator which is a product of basic fields
is the sum of the $\r$-charges of its constituents.
Neutral operators are those gauge invariant operators whose
$\r$-charge vanishes.

\subsection {Orbifold projection}

To construct the daughter theory one selects from the basic
variables of the parent theory those degrees of freedom which
are invariant under the chosen projection group \P.
The resulting projected link variables and matter fields,
viewed as $\Nc \times \Nc$ matrices, satisfy the constraints
(for each $j = 1, \ldots, p$):
\begin{eqnarray}
\label{eq:orbU}
     \U[\l]&=&
                \gamma_j\, \U[\l]\, \gamma_j^{-1}	\,,\\
     \E[\l]&=&
                \gamma_j\, \E[\l]\, \gamma_j^{-1}	\,,\\
     \phih^a[s]&=& e^{i{2\pi r_j[\phih^a]}/{k_j}} \;
                \gamma_j\, \phih^a[s]\, \gamma_j^{-1}	\,,\\
     \pih^a[s]&=& e^{i{2\pi r_j[\phih^a]}/{k_j}} \;
                \gamma_j\, \pih^a[s]\, \gamma_j^{-1}	\,,\\
     \psih^b[s]&=& e^{i{2\pi r_j[\psih^b]}/{k_j}} \;
                \gamma_j\, \psih^b[s]\, \gamma_j^{-1}	\,.
\label{eq:orbpsi}
\end{eqnarray}
If every $\Nc \times \Nc$ matrix is viewed as a collection of
$m^2$ blocks, each of which is $N \times N$, then the net effect
of the constraints (\ref {eq:orbU})--(\ref{eq:orbpsi}) is to
eliminate all but $m$ of these $N \times N$ blocks.
It is convenient to represent the surviving field content
using a graph containing $m = k_1 k_2\cdots k_p$ vertices
arranged as a discretized $p$-dimensional torus with periodicities
$k_1$, $k_2$, \ldots, $k_p$, in each direction.
Each vertex represents one of the $U(N)$ factors of the
$[U(N)]^m$ product gauge group of the daughter theory.
This graph (often called ``theory space'') will be denoted as ${\T}$,
and its vertices will be labeled by a $p$-dimensional vector $\j{\in}\T$
whose $i$'th component
takes integer values from $0$ and $k_i{-}1$ (modulo $k_i$).

Under the projection,
the parent link variables and conjugate electric fields break up into
$m$ distinct $U(N)$ link variables and electric fields,
one each associated with every vertex in theory space;
these will be denoted $\hat U^\j[\l]$ and $\hat E^\j[\l]$.
Each parent matter field breaks
up into $m$ distinct daughter fields transforming as either adjoints
or bifundamentals under the $[U(N)]^m$ product gauge group,
depending on whether the $\r$-charge of the field is zero or non-zero.
Each bifundamental matter field may be associated with a bond
in the theory space graph connecting the vertex under which
the field transforms as a fundamental with the vertex under which
it transforms as an antifundamental, while adjoint matter fields
may be associated with bonds which begin and end at the same vertex.
The daughter matter fields will be denoted
$\phih^{a\,\j}[s]$, $\psih^{b\,\j}[s]$, {\em etc.},
with $\phih^{a\,\j}[s]$, for example, represented by a bond running
from site $\j$ to site $\j+\r[\phih^a]$.
Illustrations of the resulting theory space graphs for various
examples may be found, for example, in Refs.~\cite{Schmaltz,KaplanUnsal,KUY}.

For any classical operator $\hat A^\parent$ in the parent theory,
we define its daughter theory counterpart $\hat A^\daughter$ to be
the result of replacing every variable in $\hat A^\parent$ by its
orbifold projection, as described by Eqs.~(\ref{eq:orbU})--(\ref{eq:orbpsi}).
We will write this relation as
\begin {equation}
    \hat A^\parent \to \hat A^\daughter \,.
\end {equation}
Single-trace observables in the parent,
divided by $\Nc$, map to corresponding single trace observables
in the daughter which are divided by $N$ and averaged over theory space.
For example,
\begin {equation}
    \frac 1\Nc \, \tr (\hat U[C])
    \to
    \frac 1m \, \sum_{\j\in\T} \>
    \frac 1N \, \tr (\hat U^\j[C]) \,.
\end {equation}

For this class of orbifold projections,
the daughter theory
will have a $\Z_{k_1} \times \Z_{k_2} \times \cdots \times \Z_{k_p}$
global symmetry under which different $U(N)$ factors of the product
gauge group are cyclically permuted, in a manner reflecting
the discrete translation symmetry of the periodic theory space \T.
Daughter theory operators which are invariant under these
theory space translations will be termed {\em neutral}
(while neutral parent theory operators are those which are invariant
under the projection group).
Note that the orbifold projection gives a one-to-one and onto
mapping between the set of all neutral gauge-invariant single-trace operators
in the parent theory and the set of all neutral gauge-invariant single-trace
operators in the daughter.

The fundamental operators
of the daughter theory are defined to satisfy the
commutation relations:
\begin{eqnarray}
    \left[ \U_{ij}^\j [\l], \U_{kl}^{\j'}[\l'] \right]
    &=&
    0 \,,
\label {eq:commDU}
\\
\label {eq:commDE}
    \left[ \E_{ij}^\j [\l], \U_{kl}^{\j'}[\l'] \right]
    &=&
    \frac{1}{N} \, \delta^{\j \j'} \, \delta_{\l \l'} \, \delta_{kj} \,
    \U_{il}^\j [\l] \,,
\\
    \left[ \E_{ij}^\j [\l], \U_{kl}^{\j'}[\bar\l'] \right]
    &=&
    -\frac{1}{N} \, \delta^{\j \j'} \, \delta_{\l \l'} \, \delta_{il} \,
    \U_{kj}^{\j}[\bar\l] \,,
\\
    \left[ \E_{ij}^\j [\l], \E_{kl}^{\j'}[\l'] \right]
    &=&
    \frac{1}{N} \, \delta^{\j \j'} \, \delta_{\l \l'}
    \left(
	\delta_{kj} \, \E_{il}^{\j}[\l] - \delta_{il} \, \E_{kj}^{\j}[\l]
    \right) ,
\label {eq:gaugecom}
\\
    \left[ \phih^{a\, \j}_{ij}[s], \pih^{a' \j'}_{lk}[s']^\dagger \right]
    &=&
    \left[ \phih^{a\, \j}_{ij}[s]^\dagger, \pih^{ a' \j'}_{lk}[s'] \right]
    =
    \frac{i}{N} \, \delta^{\j \j'} \, \delta_{ss'} \,
    \delta^{a a'} \, \delta_{il} \, \delta_{kj} \,,
\\
    \left\{ \psih^{a\, \j}_{ij}[s], \psih^{a'\j'}_{lk}[s']^\dagger\right\}
    &=&
   \frac{1}{N} \, \delta^{\j\j'} \, \delta_{ss'} \,
   \delta^{a a'} \, \delta_{il} \, \delta_{kj} \,.
\label {eq:commDpsi}
\end{eqnarray}
To see that this normalization of commutators
(with factors of $1/N$ instead of $1/\Nc$)
is appropriate in the daughter theory,
one may check that this choice makes normalized
traces of commutators coincide in the parent and daughter theories.
For example,
\begin {eqnarray}
    \frac 1\Nc \,
    \tr \left(\left[\phih^a[s],\pih^{a'}[s']^\dagger \right]\right)
    &\equiv&
    \frac 1\Nc \, \left[ \phih^a_{ij}[s], \pih^{a'}_{ij}[s']^\dagger \right]
    =
    i \, \delta_{ss'} \, \delta^{a a'} 
\nonumber\\
    \downarrow \qquad\qquad &&
\\
    \frac 1m \, \sum_{\j\in\T} \>
    \frac 1N \,
    \tr\left(\left[\phih^{a\,\j}[s],\pih^{a'\j}[s']^\dagger \right]\right)
    &\equiv&
    \frac 1m \, \sum_{\j\in\T} \>
    \frac 1N \,
    \left[ \phih^{a\,\j}_{ij}[s], \pih^{a'\j}_{ij}[s']^\dagger \right]
    =
    i \, \delta_{ss'} \, \delta^{a a'} \,,
\nonumber
\end {eqnarray}
and similarly for other commutators.

For the Hamiltonian, the appropriate mapping to the daughter theory
involves replacing every variable by its orbifold projection,
and then rescaling the operator by an overall factor of $N/\Nc = 1/m$.
This may equivalently be expressed as
\begin {eqnarray}
    {\hat H^\parent \over \Nc^2} &\to&
    {\hat H^\daughter \over m \, N^2} \,.
\label {eq:H-map}
\end {eqnarray}
As a shorthand,
we will write this relation as $\hat H^\parent \mapsto \hat H^{(\d)}$.
A simple check that this is the natural relation
between parent and daughter Hamiltonians is provided
by a $\Z_m$ orbifold projection applied
to $U(mN)$ pure gauge theory.
The daughter theory is just
$m$ decoupled copies of $U(N)$ pure gauge theory,
and the factor of $m$ in the
denominator of (\ref {eq:H-map})
ensures that ground state energies are correctly identified.

For our specific parent Hamiltonian
(\ref {eq:parentHamiltonian})--(\ref {eq:pHfermion}),
the mapping (\ref {eq:H-map}) produces a daughter Hamiltonian
\begin{equation}
  \H^{(\d)} \equiv
  \H^{(\d)}_{\rm gauge} + \H^{(\d)}_{\rm scalar} + \H^{(\d)}_{\rm fermion} \,,
\label {eq:daughterH}
\end{equation}
with
\begin{equation}
   \H^{(\d)}_{\rm gauge}
   =
    N
    \biggl\{ \,
	\frac{1}{4\betatilde} \,
	\sum_{\j\in \T}
	\sum_{\l\in\Lambda} \,
	\tr\, \E^\j[\l]^2
	-
 	\betatilde \,
	\sum_{\j\in \T}
	\sum_{p\in\Lambda} \,
	\tr\left(
	\U^\j[\partial p] + \U^\j[\overline{\partial p}] \right)
    \biggr\} \,.
\label {eq:dHgauge}
\end{equation}
Note the equality of 't Hooft couplings (given by $1/\betatilde$)
in parent and daughter theories.
The scalar Hamiltonian of the daughter theory is
\begin{eqnarray}
   \H^{(\d)}_{\rm scalar}
     &=&
     N
    \sum_{s\in \Lambda}
     \biggl\{ \>
    \sum_{\j \in \T}
    \tr\left(\pih^{a\,\j}[s]^\dagger \, \pih^{a\,\j}[s] \right)
    +
	\Nc \, V\!\Bigl[
	\sum_{\j\in \T}
	\smash{\frac{\tr}  {\Nc}} \!\!
	\left(\phih^{a\,\j}[s]^\dagger \phih^{a\,\j}[s]\right)
	\Bigr]
    \biggr\}
\nonumber\\ && \qquad {}
    -
    N
    \sum_{\l=\langle ss' \rangle \in\Lambda} \>
    \sum_{{\j} \in \T} \>
	\frac\kappa 2 \>
	\tr \!
	\left(
	    {\phih}^{a\,\j}[s]^\dagger \U^{\j}[\l] \,
	    \phih^{a\,\j}[s'] \, \U^{{\j+ \r_{a}}}[\bar \l]
	\right)
    \,,
\label {eq:dHscalar}
\end{eqnarray}
with $\r_a\equiv \r[\phih^a]$,
while the fermionic Hamiltonian is
\begin{eqnarray}
   \H^{(\d)}_{\rm fermion}
   &=&
    N \>
    \sum_{s\in\Lambda} \>
    \sum_{{\j} \in \T} \>
	m_{\rm f} \, \xi_s \>
	\tr \! \left(\psih^{b\,\j}[s]^\dagger \, \psih^{b\,\j}[s]\right)
\nonumber\\ && {} +
    N
    \sum_{\l=\langle ss' \rangle \in\Lambda} \>
    \sum_{{\j} \in \T} \>
    \frac{\kappa}{2i} \>
	\tr \!
	\left(
	    \psih^{b\,\j}[s] \, \eta[\l] \, \U^{\j}[\l] \,
	    \psih^{b\,\j}[s'] \, \U^{\j + \r^{b}}[\bar \l]
	\right)
    \,,
\label {eq:dHfermion}
\end{eqnarray}
with $\r^b\equiv \r[\psih^b]$.
In both matter field Hamiltonians,
an implied summation over flavor indices is
present in each trace.

\subsection {Coherence algebra and coherent states}

As for any gauge theory with adjoint matter,
the generators of the daughter theory coherence algebra $\g_{\rm daughter}$
are Wilson loops decorated with multiple insertions of matter fields
and their conjugate momenta.
The algebra $\g_{\rm daughter}$ has the same form shown in
Eq.~(\ref{eq:gsf-algebra}),
except that every variable now has an additional theory space index,
and the natural overall factor is $N$ instead of $\Nc$.
Every gauge invariant single-trace operator corresponds to a closed path
in theory space, in addition to forming a closed loop in the physical lattice.
Coefficients of individual terms now depend on a starting point in
theory space, as well as depending on the choice of loops and starting sites
in the (extended) lattice.

The appropriate mapping from elements of the parent theory coherence algebra
to those of the daughter is the same as the relation between Hamiltonians,
namely,
\begin {equation}
    {\hat \Lambda^\parent \over \Nc^2} \to
    {\hat \Lambda^\daughter \over m \, N^2} \>,
\label {eq:coher-proj}
\end {equation}
or $\hat \Lambda^\parent \mapsto \hat\Lambda^\daughter$.
However,
the coherence algebra in the daughter theory is \emph{not} just the result of
applying the orbifold projection to the parent coherence algebra
(\ref{eq:gsf-algebra});
due to the product gauge group structure,
$\g_{\rm daughter}$ is much larger than the image of $\g_{\rm parent}$
under the mapping (\ref {eq:coher-proj}).
For example, $\g_{\rm daughter}$
contains independent (anti-Hermitian) linear combinations
of Wilson loops for each one of the $m$ different gauge group factors,
instead of the single linear combination appearing in the parent algebra.

Nevertheless, there is a simple relation between the parent
and daughter coherence algebras:
the orbifold projection maps the neutral subalgebra
$\h_{\rm parent}$ of the parent theory
to the subalgebra $\h_{\rm daughter}$ of the coherence
algebra $\g_{\rm daughter}$ consisting of all neutral generators.
Somewhat more explicitly, $\h_{\rm daughter}$ consists of
all anti-Hermitian operators of the form
\begin{eqnarray}
    \hat\Lambda^\daughter(a,b,c)
    &\equiv &
    N \, \sum_{\j\in\T} \biggl\{
      \sum_{C\subset\bar\Lambda} a^C \, \tr (\U^\j[C])
    + \sum_{C\subset\Lambda} \sum_{\l\subset C} \>
	b^{\l C} \, \tr({:}\E^\j[\l]\,\U^\j[C]{:})
\label {eq:orbifold-algebra}
\\
    &+&
    \sum_{C\subset\bar\Lambda_{\rm s}} \, \sum_{s\subset C} \>
    \left[ \,
	c_a^{s C} \, \tr({:}\pih^{a\,\j}[s]\,\U^{\j+\r_a}[C]{:})
	- (c_a^{s C})^* \,
	\tr ({:}\pih^{a\,\j}[s]^\dagger\,\U^{\j+\r_a}[\bar C]{:})
    \right]
    \biggr\} \,,
\nonumber
\end{eqnarray}
where $\hat U^\j[C]$ represents the ordered product of variables
around a closed loop $C$ in the original or extended lattice,
as indicated, with the first variable associated with point $\j$
in theory space and theory space labels of all subsequent variables
uniquely dictated by gauge invariance.%
\footnote
    {
    When the daughter theory contains bifundamental representation
    matter fields, not all closed loops on the extended lattice
    correspond to gauge-invariant operators;
    only those loops which may also be
    associated with closed paths in theory space represent gauge
    invariant operators.
    The loop sums in the algebra (\ref {eq:orbifold-algebra})
    should be understood as only including these loops.
    }
In the above expression, traces are over $N{\times}N$ matrices,
and the overall factor of $N$ ensures that structure constants of
the subalgebra ${\h}_{\rm daughter}$ are independent of $N$.

As was done in the parent theory,
we will choose the base state
of the daughter theory
to be a product,
$
    |0\rangle =
    |0\rangle_{\rm gauge} \otimes
    |0\rangle_{\rm scalar} \otimes
    |0\rangle_{\rm fermion}
$,
with $|0\rangle_{\rm gauge}$
the strong-coupling pure gauge ground state
(whose configuration space wavefunction is unity),
and $|0\rangle_{\rm scalar}$ the decoupled product of Gaussian
states annihilated by $\phih^{a\,\j}[s] + i \pih^{a\,\j}[s]$ and
$\phih^{a\,\j}[s]^\dagger +i \pih^{a\,\j}[s]^\dagger$
(with wavefunction
$
    \langle \phi|0\rangle_{\rm scalar}
    =
    e^{-N \sum_{s\in\Lambda}
	  \sum_{\j\in \T}
	  \sum_a
	  \tr (\phi^{a\,\j}[s] \, \phi^{a\,\j}[s]^*)}
$).
The fermion state $|0\rangle_{\rm fermion}$ will be the
ground state of the fermionic mass term in (\ref {eq:dHfermion}),
which is annihilated by $\psih^{b\,\j}[s]$ on even sites and by
$\psih^{b\,\j}[s]^\dagger$ on odd sites
(so it contains $m \, N^2 \Nf$ fermions at every odd site,
and no fermions at even sites).
Once again, the resulting base state is gauge invariant.
Large-$N$ expectation values of all decorated Wilson loops
vanish in this state,
except for ``loops'' which are the trace of a product of matter fields
at a single site.

Coherent states in the daughter theory are defined, as usual,
by the action of the coherence group $\G_{\rm daughter}$ acting
on the base state $|0\rangle$.
The same coherent state properties discussed earlier for
the parent theory apply equally well to daughter theory
coherent states:
they provide an overcomplete basis in the Hilbert space of
gauge invariant states,
any operator is uniquely defined by its coherent state expectation values,
and a coadjoint orbit of the coherence group $\G_{\rm daughter}$
defines the classical phase space of $N=\infty$ dynamics.
The only difference from the previous results (\ref{eq:A})--(\ref{eq:hcl})
concerning the classical nature of the large $N$ limit is in the scaling of
commutators.
Given the relation (\ref {eq:coher-proj}),
it is natural to define the Poisson bracket in the daughter theory
so that
\begin{equation}
   \lim_{N\to\infty} i \, m N^2 \, \langle u|[\hat A,\hat B]|u\rangle=
        \{a(\zeta),b(\zeta)\}_{PB} \,.
\label{eq:daughter[A,B]}
\end{equation}
Consequently, the large $N$ classical Hamiltonian for the daughter theory is
\begin {equation}
   h^\daughter_{\rm cl}(\zeta) \equiv \lim_{N\to\infty} \frac{1}{m N^2} \>
                  \langle u|\H^\daughter|u\rangle \,.
\label {eq:dhcl}
\end {equation}
The neutral subalgebra $\h_{\rm daughter}$ generates the
neutral subgroup ${\bf H}_{\rm daughter}$ of the full coherence group,
and this subgroup generates the subspace of coherent states
(and hence of the classical phase space) which is invariant
under the theory space translation symmetry.

\section{Large \boldmath$N$ equivalence}

\subsection{Isomorphism of neutral coherence subalgebras}

For an element $\hat\Lambda^\parent(a,b,c)$
of the coherence algebra $\g_{\rm parent}$ to be neutral
(invariant under the projection group \P), it must include only single traces
composed of products of operators whose $\r$-charges sum to zero.
This is exactly the same condition as the requirement that
each single-trace operator in $\hat\Lambda^\parent(a,b,c)$,
after orbifold projection,
yield an operator representable as a closed loop in theory space.
Neutrality of $\hat\Lambda^\parent(a,b,c)$
implies that it is mapped to a non-zero gauge-invariant operator
in the daughter theory, which is necessarily invariant
under theory space translations.
(In contrast, non-neutral operators in the parent theory
map to zero under the orbifold projection.)
With our parameterization (\ref {eq:orbifold-algebra})
of the neutral coherence algebra of the daughter theory,
every neutral element $\hat\Lambda^\parent(a,b,c)$ maps
precisely to $\hat\Lambda^\daughter(a,b,c)$.
Hence, the orbifold mapping $\h_{\rm parent} \mapsto \h_{\rm daughter}$
is one-to-one and onto.

An essential point is that this mapping also preserves
the Lie algebra structure, so this mapping is an isomorphism
between $\h_{\rm parent}$ and $\h_{\rm daughter}$.
In other words, if
\begin{equation}
     \Bigl[\hat\Lambda^\parent(a_1,b_1,c_1),\,
     \hat\Lambda^\parent(a_2,b_2,c_2) \Bigr] =
     \hat\Lambda^\parent(a_{12},b_{12},c_{12}) \,,
\end{equation}
then
\begin{equation}
     \Bigl[\hat\Lambda^{(\d)}(a_1,b_1,c_1),\,
     \hat\Lambda^{(\d)}(a_2,b_2,c_2) \Bigr] =
     \hat\Lambda^{(\d)}(a_{12},b_{12},c_{12}) \,,
\end{equation}
where the common parameters
$a_{12},b_{12},c_{12}$ of both results depend on
$a_{1},a_{2}$ \emph{etc.}, but are independent of $\Nc$ and $N$.
Their explicit form is somewhat lengthy; what is important is that for
given input parameters, the result parameters are the same in both algebras.

It may be instructive to illustrate this with a simple example.
Let $C$ be a loop on the spatial lattice $\Lambda$
which is composed of a path $\Gamma_1$ running from site $s_1$ to site $s_2$,
followed by a path $\Gamma_2$ running from $s_2$ back to $s_1$.
And similarly, let $C' = \Gamma_1' \Gamma_2'$ be a loop which
starts at site $s_1'$ and passes through site $s_2'$.
Consider the following two generators in ${\bf h}_{\rm parent}$:
\begin{eqnarray}
    \hat\Lambda^\parent_{1} 
    &=&
    \Nc \; \tr
    \Bigl(
	\phi^{a}[s_1] \, U[\Gamma_1] \, \phi^{a}[s_2]^\dagger \, U[\Gamma_2]
    \Bigr) \,,
\\
    \hat\Lambda^\parent_{2} 
    &=&
    \Nc \; \tr \Bigl(
    \pi^{a'}[s'_1]^\dagger \, U[\Gamma'_1] \, \phi^{a'} [s'_2] \, U[\Gamma'_2]
    \Bigr) \,.
\end{eqnarray}
The commutator of these generators is
\begin{eqnarray}
     \Bigl[ \hat\Lambda^\parent_{1}, \, \hat\Lambda^\parent_{2} \Bigr]
     &=&
     \delta^{aa'} \, \delta_{s_1 s'_1} \;
     \Nc \;\tr
     \Bigl(
	 \phi^{a}[s_2]^\dagger \,U[\Gamma_2 \Gamma'_1] \,
	 \phi^{a} [s'_2] \, U[\Gamma'_2 \Gamma_1]
     \Bigr)
\nonumber \\
     &\equiv&
     \delta^{aa'} \, \delta_{s_1 s_1'}\;\hat\Lambda^\parent_{3} \,.
\end{eqnarray}
The corresponding generators in ${\bf h}_{\rm daughter}$ are
\begin{eqnarray}
    \hat\Lambda^\parent_{1}
    \mapsto
    \hat\Lambda_{1}^{(\d)}
    &=&
    N  \, \sum_{\j \in \T} \, \tr \;
    \Bigl(
	\phi^{a\,\j}[s_1] \, U^{\j + {\bf r}_a}[\Gamma_1] \,
	\phi^{a\,\j}[s_2]^\dagger \, U^{\j}[\Gamma_2]
    \Bigr) \,,
\\
    \hat\Lambda^\parent_{2}
    \mapsto
    \hat\Lambda_{2}^{(\d)}
    &=&
    N \, \sum_{\j' \in \T} \, \tr \;
    \Bigl(
	\pi^{a'\j'}[s'_1]^\dagger \, U^{\j'}[\Gamma'_1] \,
	\phi^{a'\j'}[s'_2] \, U^{\j' + {\bf r}_{a'}}[\Gamma'_2]
    \Bigr) \,,
\end{eqnarray}
and their commutator, in the daughter theory, is
\begin{eqnarray}
     \Bigl[\hat\Lambda_{1}^{(\d)}, \, \hat\Lambda_{2}^{(\d)}\Bigr]
     &=&
     \delta^{aa'} \, \delta_{s_1 s'_1} \, N \,
     \sum_{\j \in \T}
     \tr \Bigl(
	 \phi^{a\, \j}[s_2]^\dagger
	 \,U^{\j}[\Gamma_2\Gamma'_1] \, \phi^{a\, \j}[s'_2] \,
	 U^{\j+{\bf r}_{a}} [\Gamma'_2\Gamma_1]
     \Bigr)
\nonumber \\
     &\equiv&
     \delta^{aa'} \, \delta_{s_1 s_1'}\;\hat\Lambda_{3}^{(\d)} \,.
\end{eqnarray}
The result $\hat\Lambda^\daughter_3$ coincides, as claimed,
with the image of $\hat\Lambda^\parent_3$ under the orbifold mapping,
$\hat\Lambda^\parent_3 \mapsto \hat\Lambda^\daughter_3$.
One may verify that this is true in general.

\subsection {Equivalence of large $N$ observables}

The base states, in both parent and daughter theories,
were chosen to make the evaluation of expectation values in these
states trivial.
In both theories, the only physical observables
which have non-vanishing large-$N$ base state expectation values are
(products of) traces of products of matter fields at a single site.
And the normalization of commutators in the two theories,
Eqs.~(\ref {eq:commPU})--(\ref{eq:commPpsi}) and
(\ref{eq:commDU})--(\ref{eq:commDpsi}),
are exactly what is required so that the non-zero expectation
values of corresponding observables coincide in the large $N$ limit.
For example,
\begin {eqnarray}
    \langle 0|
	\frac 1\Nc \, \tr \, \phih^a[s]^\dagger \phih^{a'}[s']
    |0\rangle_{\rm p}
    =
    \frac 1m \sum_{\j\in\T} \>
    \langle 0|
	\frac 1N \, \tr \, \phih^{a\,\j}[s]^\dagger \phih^{a'\j}[s']
    |0\rangle_{\rm d}
    &=&
    \half \, \delta^{aa'} \delta_{ss'} \,,
\\
    \langle 0|
	\frac 1\Nc \, \tr \, \psih^b[s]^\dagger \psih^{b'}[s']
    |0\rangle_{\rm p}
    =
    \frac 1m \sum_{\j\in\T} \,
    \langle 0|
	\frac 1N \, \tr \, \psih^{b\,\j}[s]^\dagger \psih^{b'\j}[s']
    |0\rangle_{\rm d}
    &=&
    \half \, \delta^{bb'} \delta_{ss'} (1 {-} \xi_s) \,,
\end {eqnarray}
where $\langle 0| \cdots |0\rangle_{\rm p}$
and $\langle 0| \cdots |0\rangle_{\rm d}$
denote expectation values in the indicated theories.
One may easily check that this correspondence is true in general.

Since the classical phase space is a coadjoint orbit of
the coherence group, with the particular coadjoint orbit
determined by base state expectation values \cite {LGY-largeN},
the isomorphism between the neutral coherence subgroups
$\h_{\rm parent}$ and $\h_{\rm daughter}$,
combined with the isomorphism between large-$N$ base state
expectation values of corresponding observables,
immediately implies that the neutral sectors of the large-$N$
phase spaces of parent and daughter theories are isomorphic.

This also implies that corresponding classical observables
in parent and daughter theories have coinciding values
throughout the neutral sector of the large $N$ phase space.
To see this, consider how observables
change along geodesics in phase space, which are images
of geodesics (or one-parameter subgroups) of the coherence group.
Any specific coherent state $|u\rangle$ may be connected
with the base state $|0\rangle$ by the action of some one-parameter
subgroup of the coherence group,
\begin {equation}
    |u\rangle = |\Lambda,1\rangle \,, \qquad \hbox{with} \qquad
   |\Lambda,t\rangle \equiv e^{t \hat\Lambda} \, |0\rangle \,.
\end {equation}
The expectation value of any operator $\hat {\cal O}$
changes along the geodesic according to the equation
\begin{equation}
    \frac{d}{dt} \, \langle \Lambda,t|\,\hat {\cal O}\,|\Lambda,t\rangle =
    \langle \Lambda,t|\,[\hat {\cal O}, \hat\Lambda]\,|\Lambda,t\rangle \,.
\label{eq:geodesic1}
\end{equation}
This shows that expectation values of
elements of the coherence algebra obey a closed set
of first-order differential equations.
More generally, if ${\cal S}_K = \{\hat {\cal O}_\alpha\}$ is the set of
all neutral single-trace classical observables containing at most $K$
insertions of bosonic conjugate momenta (scalar or gauge),
then, for any value of $K$,
expectation values of this set of observables
obey a closed set of first order equations,
\begin{equation}
    \frac{d}{dt} \,
    \langle \Lambda,t|\, \hat {\cal O}_\alpha \,|\Lambda,t\rangle =
    {\cal M}_{\alpha\beta} \,
    \langle \Lambda,t|\, \hat {\cal O}_\beta \,|\Lambda,t\rangle \,,
\label{eq:geodesic2}
\end{equation}
with coefficients ${\cal M}_{\alpha\beta}$ which may be computed
using the canonical commutation relations,
and which are $N$-independent.

The essential point is that these ``geodesic'' equations
are the {\em same} in the parent and daughter theories,
provided $\hat\Lambda$ is a generator in the neutral subspace.
If $\hat\Lambda^\parent \mapsto \hat\Lambda^\daughter$
and $\hat {\cal O}^\parent_\alpha \to \hat {\cal O}^\daughter_\alpha$
are corresponding generators and observables in the parent and
daughter theories, then (using the canonical commutation relations),
one may verify that
\begin{equation}
    [\hat O^\parent_{\alpha}, \hat\Lambda^\parent] \to
    [\hat O^\daughter_{\alpha}, \hat\Lambda^\daughter] \,.
\label{eq:same-commutators}
\end{equation}
This implies that the coefficients ${\cal M}_{\alpha\beta}$
are the same in parent and daughter theories.

Combined with the large-$N$ equivalence of base state expectation
values, this means that the geodesic equations (\ref{eq:geodesic2})
within the neutral sector,
as well as their initial values,
coincide in parent and daughter theories.
Consequently, the solutions of these geodesic equations (at $N=\infty$)
also coincide, showing that corresponding physical operators have
identical values,
\begin{equation}
    \lim_{\Nc\to\infty}
    \langle u|\hat {\cal O}^\parent_\alpha|u\rangle_{\rm p}
    =
    \lim_{N\to\infty}
    \langle u|\hat {\cal O}^\daughter_\alpha|u\rangle_{\rm d} \,,
\label {eq:observable-equiv}
\end{equation}
throughout the neutral sector of the large-$N$ phase space.
In other words, the classical phase space observables 
associated with corresponding physical operators are identical
in the neutral sectors of the parent and daughter large-$N$ phase spaces.

\subsection{Equivalence of large $N$ dynamics}

The equivalence (\ref {eq:observable-equiv}) between
corresponding observables (in the neutral sector),
combined with relation (\ref {eq:H-map}) between parent and daughter
quantum Hamiltonians
and the definitions (\ref {eq:phcl}) and (\ref {eq:dhcl})
of the large-$N$ classical Hamiltonians
in parent and daughter theories,
immediately yields the essential result that the
large-$N$ classical Hamiltonians
coincide on the neutral sectors of their respective phase spaces,
\begin {equation}
    h^\parent_{\rm cl}(\zeta) = h^\daughter_{\rm cl}(\zeta) \,.
\end {equation}
The isomorphism between the neutral coherence subgroups of the 
parent and daughter theories directly implies that the symplectic structures
of the parent and daughter large $N$ phase spaces are identical within
their neutral sectors.
Therefore, all dynamics of these large $N$ classical systems
coincide (within their neutral sectors).

This means that any physical quantity which can be extracted from
the large $N$ classical dynamics, in the neutral sector, will
coincide between parent and daughter theories.
If the symmetries defining the neutral sectors
(the orbifold projection symmetry in the parent,
and theory space translation symmetry in the daughter)
are {\em not} spontaneously broken
for some chosen values of the coupling constants of the theories,
then the minimum of the classical Hamiltonian will lie
in the neutral sector of the large $N$ phase space for both theories.
In this case, not only will the ground state energy and
expectation values of corresponding single trace observables
coincide in the two theories,
so will small oscillation frequencies for deformations away from
the minimum which lie in the neutral sector.
Such small oscillation frequencies are the large-$N$ limits
of excitation energies in the underlying quantum theories,
for states which can be produced by acting with single trace
operators on the vacuum.
Therefore, the large $N$ limit of the particle spectrum of the parent and
daughter theories will coincide in all neutral symmetry channels.
Two-body decay amplitudes of such particles scale as $1/N$ as $N\to\infty$,
while $p \leftrightarrow q$ particle scattering amplitudes
scale as $1/N^{p+q-2}$.
The leading large-$N$ behavior of such amplitudes is completely
determined by the large $N$ classical dynamics
(in exactly the same way that the usual $\hbar\to0$ classical action
determines all tree-level diagrams).
Consequently, scattering amplitudes of particles
are related between parent and daughter theories.
Specifically,
if $\Gamma_{p,q}$ denotes an $p\leftrightarrow q$ particle scattering amplitude,
involving corresponding particles in the neutral sectors of each theory,
then
\begin {equation}
    \lim_{\Nc\to\infty} \> (\Nc^2)^{(p+q-2)/2} \> \Gamma^\parent_{p,q}
    =
    \lim_{N\to\infty} \> (m N^2)^{(p+q-2)/2} \> \Gamma^\daughter_{p,q} \,.
\end {equation}
The equivalent relation for
connected correlators of corresponding classical operators is
\begin {eqnarray}
    \lim_{\Nc\to\infty} \> (\Nc^2)^{K-1} \>
    \langle \hat {\cal O}^\parent_1 \cdots \hat {\cal O}^\parent_K
    \rangle_{\rm conn}
    =
    \lim_{N\to\infty} \> (m N^2)^{K-1} \>
    \langle \hat {\cal O}^\daughter_1 \cdots \hat {\cal O}^\daughter_K
    \rangle_{\rm conn} \,.
\end {eqnarray}

Alternatively, if the symmetries defining the neutral sectors
{\em are} spontaneously broken
(for some chosen values of coupling constants),
then the equivalence between
the large $N$ dynamics within the neutral sectors of the two theories
(which is still valid)
does not imply any equivalence between ground state energies,
correlators, or particle spectra in the two theories,
since the minimum of one or both of the large-$N$ classical
Hamiltonians no longer lie in the neutral sector.

In our lattice-regularized theories,
all symmetries are guaranteed to be unbroken in the phase
of each theory which is continuously connected to strong coupling
and large mass ({\em i.e.}, small $\betatilde$ and small $\kappa$).
But whether a given theory has a phase where the requisite symmetries
are unbroken, within which one may take a continuum limit,
will depend on the specific choice of theory.


\acknowledgments

Josh Erlich, Herbert Neuberger, and Matt Strassler are thanked for
helpful comments.
The work of P.K. and L.G.Y. is supported, in part, by the U.S. Department
of Energy under Grant No.~DE-FG03-96ER40956;
the work of M.\"U. is supported by DOE grant  DE-FG03-00ER41132.

\sloppy

\end {document}